\documentclass{aa}  
\usepackage[utf8]{inputenc}

\usepackage{graphicx}
\usepackage{siunitx}
\usepackage{amsmath}
\usepackage{textcomp}
\usepackage{multirow}                
\usepackage{booktabs}                              
\usepackage{multicol,lipsum}
\usepackage{colortbl}

\usepackage{txfonts}
\usepackage[]{hyperref}
\usepackage{hyperref}
 \hypersetup{
     colorlinks=true,
     linkcolor=blue,
     filecolor=blue,
     citecolor =blue,      
     urlcolor=blue,
     }

\DeclareRobustCommand{\ion}[2]{%
\relax\ifmmode
\ifx\testbx\f@series
{\mathbf{#1\,\mathsc{#2}}}\else
{\mathrm{#1\,\mathsc{#2}}}\fi
\else\textup{#1\,{\mdseries\textsc{#2}}}%
\fi}

\usepackage{comment}
\usepackage{balance}

\begin{document}

   \title{The GAPS programme at TNG XXVI. Magnetic activity in M stars: spectroscopic monitoring of AD Leonis}

   \author{C. Di Maio
          \inst{1,2}
          \and
          C. Argiroffi\inst{1,2}
          \and
          G. Micela\inst{2}
          \and
          S. Benatti\inst{2}
          \and A.F. Lanza\inst{3}
          \and G. Scandariato\inst{3}
          \and J. Maldonado\inst{2}
          \and A. Maggio\inst{2}
          \and L. Affer\inst{2}
          \and R. Claudi\inst{4}
          } 

   \institute{Università degli Studi di Palermo, Dipartimento               di Fisica e Chimica, via Archirafi 36, Palermo,               Italy.\\
              \email{claudia.dimaio@inaf.it}
         \and
             INAF – Osservatorio Astronomico di Palermo, Piazza del Parlamento, 1, 90134 Palermo, Italy
        \and
            INAF – Osservatorio Astrofisico di Catania, Via S. Sofia 78, 95123 Catania, Italy
        \and 
            INAF – Osservatorio Astronomico di Padova, Vicolo dell’Osservatorio 5, 35122 Padova, Italy}

   \date{Received; Accepted}

\titlerunning{Spectroscopic monitoring of AD Leonis}
\authorrunning{Di Maio et al.}
 
  \abstract
   {Understanding stellar activity in M dwarfs is fundamental to improving our knowledge of the physics of stellar atmospheres and for planet search programmes. High levels of stellar activity (also frequently associated with flare events) can cause additional variations in the stellar emission that contaminate the signal induced by a planet and that need to be corrected. The study of activity indicators in active stars can improve our capability of modelling the signal generated by magnetic activity.}
  {In this work we present measurements of activity indicators at visible wavelength for a star with a high activity level, AD Leonis, observed with HARPS in 2006, and HARPS-N in 2018. Our aim is to understand the behaviour of stellar chromospheres of M stars, studying the more sensitive chromospheric activity indicators. We also focus on characterising their variability and on finding the correlations among these indicators to obtain information on the origin of the magnetic activity in low-mass stars.}
  {We performed a study of the main optical activity indicators (\ion{Ca}{ii} H\&K, Balmer lines, \ion{Na}{i} D$_{1,2}$ doublet, \ion{He}{i} D$_3$, and other helium lines) measured for AD Leonis using the data provided by the HARPS-N high-resolution spectrograph at the Telescopio Nazionale Galileo in 2018, and by the HARPS instrument at La Silla observatory in 2006. Spectra were flux-calibrated in units of flux at the stellar surface. We measured excess flux of the selected activity indicators. The correlations between the different activity indicators as well as the temporal evolution of fluxes were analysed.
  A stellar flare was identified during the 2018 observing run and the H$\alpha$, H$\beta$, \ion{He}{i} 4471 \AA ,\ and \ion{He}{i} 5876 \AA \ lines were analysed in detail by fitting the line profiles with two Gaussian components.  }
  {We found that the \ion{Ca}{ii} H\&K flux excesses are strongly correlated with each other, but the \ion{Ca}{ii} H\&K doublet is generally less correlated with the other indicators. Moreover, H$\alpha$ is correlated with \ion{Na}{i} doublet and helium lines.
  Analysing the time variability of flux of the studied lines, we found a higher level of activity of the star during the observations in 2018 than in 2006, while \ion{Ca}{ii} H\&K showed more intense emission on spectra obtained during the observations in 2006. 
  Thanks to a detailed analysis of selected line profiles, we investigated the flare evaluating the mass motion during the event. }
   {}


   \maketitle
%
\section{Introduction}
    Magnetic activity in late-type main-sequence stars is  observable evidence of the stellar magnetic fields. The generation and intensification of surface magnetic fields in solar stars are generally due to a complex dynamo mechanism, whose efficiency is determined by the interaction between differential rotation and subphotospheric convection into the stellar interior and in which meridional circulation plays an important role \citep{2015SSRv..196..303B,2017LRSP...14....4B,2020LRSP...17....4C}. 
    Magnetic fields reach the stellar surface and manifest themselves in a variety of phenomena that we call stellar activity: starspots, chromospheric plages, heating of the chromosphere and corona, impulsive flares. Starspots are  a manifestation of magnetic field lines going through the stellar photosphere and obstructing the convective welling up of hot plasma, producing these cool spots that are darker than the surrounding photosphere. Chromospheric plage regions correspond to enhanced network magnetic field and facula regions in the photosphere, which might surround sunspots,  but are not necessarily associated with them. Heating of the stellar chromosphere and corona  generates chromospheric emission lines. Impulsive flares are visible in all regions of the spectrum, and are due to the reconnection of magnetic field lines \citep{1975ApJ...200..747S,1989ApJ...337..964S,2006RPPh...69..563S,2017SCPMA..60a9601C,2018ApJS..236....7H}. 
    
    M stars are small cool main-sequence stars with effective temperatures in the range 2400 - 3800 K and radii between 0.10 and 0.63 R$_\odot$; they represent 75\% of the stars in the solar neighbourhood \citep{2002AJ....124.2721R, 2006AJ....132.2360H}. They are known to generate the strongest photospheric magnetic fields among main-sequence stars \citep{1985ApJ...299L..47S, 2009ApJ...692..538R,2017NatAs...1E.184S}, showing magnetic activity as spots, flares, plages, and other brightness inhomogeneities.

    In recent years the exoplanet community have started to monitor samples of M dwarfs, aiming to search for habitable planets around these stars.
    From an observational point of view, there are more chances of finding an Earth-like planet in the habitable zone as the host star's mass decreases. Therefore, M dwarfs are extremely interesting targets for planet discovery \citep{2012A&A...541A...9G}. However, magnetic activity increases with decreasing stellar mass  \citep{1996AJ....112.2799H,2008AJ....135..785W,2017ApJ...834...85N}. 
    
    Stellar activity has effects on the search of exoplanets: in some cases the radial velocity periodicity, induced by stellar activity and rotation, may produce spurious signals that mimic planetary signals.  This was the case, for example, of AD Leonis, for which \citet{2018AJ....155..192T} proposed the existence of a planet, while \citet{2013A&A...552A.103R} and \citet{2013A&A...549A.109B} have interpreted the RV signal present in the AD Leo spectra as being due to magnetic activity;  this thesis has also been recently confirmed by \citet{2020A&A...638A...5C}. They use a multiwavelength approach (visible and near-infrared) to shown that the signal is of stellar origin. Therefore, a detailed study of magnetic activity in active M stars could improve our capability of modelling the signal generated by magnetic activity and increase our possibilities of finding new exoplanet candidates.

    In addition, stars with high levels of magnetic activity show flares more frequently than inactive stars \citep{2009AJ....138..633K}. The large amounts of energy released by flares could potentially affect the structure and temperature regime of exoplanetary atmospheres, thereby affecting the size of the habitable zone \citep{2007AsBio...7..185L}.  It is therefore crucial to better understand and quantify the activity of M dwarfs in terms of strength and variability. 
    
    Chromospheric activity is usually observed in the cores of the \ion{Ca}{ii} H\&K lines and the \ion{H}{i} Balmer lines. Other common optical activity indicators include lines such as the Na D$_{1,2}$ doublet, the \ion{Mg}{i} b triplet, or the \ion{Ca}{ii} infrared triplet. 
    A simultaneous analysis of the different indicators of magnetic activity could   increase our knowledge of the chromospheric structure and the radial-velocity variations  \citep[e.g.][]{2000A&AS..146..103M, 2013A&A...558A.141S, 2017A&A...598A..27M,2018A&A...616A.155L,2019A&A...627A.118M}. The common approach is to study the relationship between pairs of fluxes of different lines.
    
    In this paper we aim to understand the behaviour of stellar chromospheres for M stars with high levels of activity. 
    To this end, we focus our study on one M dwarf, AD Leonis, a very close active star, which was analysed through spectroscopic monitoring in the optical band. We present an analysis of fluxes and profiles of the main optical activity indicators such as chromospheric lines of \ion{H}{i}, \ion{He}{i}, \ion{Na}{i}, and \ion{Ca}{ii}.

    This paper is organised as follows. We describe the target in Sect. \ref{sec:adleo} and the observations in Sect. \ref{sec:obs}. We detail our procedure in Sect. \ref{sec:analysis}. Section \ref{sec:flux-flux} presents the analysis of the different spectral lines sensitive to the activity. A flare analysis is discussed in Sect. \ref{section:flare}. Our conclusions follow in Sect. \ref{sec:summary}.

    \section{AD Leonis}\label{sec:adleo}
    AD Leonis (AD Leo, GJ 388, BD +20 2465) is classified as dM4.5e \citep{2018AJ....155..192T} and is located in the immediate solar neighbourhood, at a distance of $\sim 4.97$ pc \citep{2018A&A...616A...1G}. \citet{2012ApJ...758...56S} estimated a radial velocity of 12.5 $\pm$ 0.2 km s$^{-1}$. 
    \citet{2013A&A...549A.109B} estimated a mass of 0.42 M$_\odot$ and a luminosity of 0.023 L$_\odot$. 
    The star has a radius of 0.436 $\pm$ 0.049 R$_\odot$ and effective temperature of 3414 $\pm$ 100 K \citep{2016ApJ...822...97H}. \citet{2012A&A...538A..25N} estimated the metallicity of AD Leo to be [Fe/H] = 0.07, while \citet{2012ApJ...748...93R} gave a value of 0.28 $\pm$ 0.17. 
    
    Based on spectropolarimetry, \citet{2008MNRAS.390..567M} reported a stellar rotation period of 2.2399 $\pm$ 0.0006 days; they also gave alternative solutions at periods of 2.2264 and 2.2537 days.
    The strongest evidence in favour of the short rotation period of AD Leo comes from the Microvariability and Oscillations of Stars (MOST) photometric observations. MOST observations were reported to contain strong evidence for a periodicity of 2.23$^{+0.36}_{-0.27}$ days \citep{2012PASP..124..545H} caused by `spots distributed at different longitudes or, possibly, that the modulation is caused by varying surface coverage of a large polar spot or a spot that is viewed nearly pole on'. This suggests a young age, estimated to be 25-300 Myr by \citet{2009ApJ...699..649S}.
    
    \citet{2016ApJ...822...97H} reported a value for $v$ sin $i$ of AD Leo equal to 2.63 km s$^{-1}$ that produced a projected rotation period of 8.38$^{+1.2}_{-1.1}$ days. Thus, since the rotation period of the star is 2.23 days, the star is oriented nearly pole-on with an inclination of $\sim 15$ degrees, confirming the value reported by \citet{2008MNRAS.390..567M} and \citet{2012AJ....143...93R}.
    
    AD Leo has been observed to be variable on longer timescales as well. \citet{2014ApJ...781L...9B} reported an approximately 7 yr activity cycle based on ASAS optical photometry and CASLEO spectroscopy. Even though the period reported in the ASAS photometry has a rather modest statistical significance with a false alarm probability (FAP)  of the order of 8\%, together with the spectroscopic data it indicates the presence of  an approximately  seven-year activity cycle in a convincing manner. 
    
    AD Leo hosts a magnetic field with properties similar to those observed for fully convective stars \citep{2008MNRAS.390..567M}.
    A high-resolution infrared spectrum of AD Leo, obtained with the Kitt Peak 4 m Fourier Transform Spectrometer, clearly shows the presence of strong magnetic fields \citep{1985ApJ...299L..47S}.
    \citet{2018MNRAS.479.4836L} inspected circularly polarised spectra and estimate an average large-scale magnetic field of $\sim 300 - 330$ G. Line broadenings in unpolarised spectra, also determined  by small-scale field structures, reveal instead  a stronger  overall magnetic field ($3100$ G, \citealt{2017NatAs...1E.184S}).

    Since AD Leo is a magnetically active star, its emission from the upper layers of the atmosphere (chromosphere and corona) is intense. In particular, in the optical band AD Leo is characterised by H$\alpha$, H$\beta$, and \ion{Ca}{ii} H\&K lines in emission, with variable line profiles (shape and intensity) that depend on the activity level at the time of observation, and by the presence of phenomena directly related to stellar magnetic activity such as flares.
    It is well known for its frequent \citep{1984ApJS...54..375P, 2006AJ....132.2360H} and strong flares \citep[e.g.][]{1991ApJ...378..725H} that have been observed and studied in the optical, extreme UV, and X-ray wavelength ranges \citep[e.g.][]{1995ApJ...453..464H, 1996A&A...310..245M, 2000A&A...354.1021F, 2003ApJ...597..535H, 2003A&A...411..587V}.

    \begin{table}[h!]
        \caption{Target characteristics}   
        \label{table:target} 
        \centering             
        \begin{tabular}{cc}   
            \toprule[0.05cm]
            \toprule
            \multicolumn{2}{c}{AD Leonis} \\
            \midrule            
            \smallskip
            Spectral type \ \tablefootmark{(a)} & M4.5e \\ 
            \enskip
            $\mathrm{M_{\star}}\mathrm{(M_{\odot})}$ \ \tablefootmark{(b)} & $\sim 0.42$\\
            \enskip
            $\mathrm{R_{\star}}\mathrm{(R_{\odot})}$ \ \tablefootmark{(c)} & $0.436 \pm 0.049$ \\
            \enskip
                    $\mathrm{log \ g}$ & $\sim 4.8$ \\
                    \enskip
                    d (pc)\ \tablefootmark{(d)} & $\sim 4.9660 \pm 0.0017$ \\\
                    \enskip
                    $\mathrm{L_{\star}}\mathrm{(L_{\odot})}$ \ \tablefootmark{(b)} & $\sim 0.023$ \\
                    \enskip
                    $\mathrm{T_{eff}}$ (K) \ \tablefootmark{(c)} & $3414 \pm 100$ \\
                    \enskip
                    $v$  sin $i$ $\mathrm{(km \ s^{-1})}$ \ \tablefootmark{(c)}& $\sim 2.63$ \\
                    \enskip
                    $\mathrm{P_{phot}}$ (d)\ \tablefootmark{(e)} & $\sim 2.23$ \\
                    \enskip
                    $\mathrm{[Fe/H]}$ \ \tablefootmark{(f)}& $0.28 \pm 0.17$ \\
                    \enskip
                    RV $\mathrm{(km \ s^{-1})}$ \ \tablefootmark{(g)}& $12.5 \pm 0.2$\\ 
                    \enskip
            B$_{pol}$ (G) \ \tablefootmark{(h)} & $\sim 300 - 330$\\
            B$_{unpol}$ (G) \ \tablefootmark{(i)} & $\sim 3100$\\
            \bottomrule[0.05cm]                
            \end{tabular}
            \tablefoot{\tablefoottext{a}{\citet{2018AJ....155..192T}} \tablefoottext{b}{\citet{2013A&A...549A.109B}} \tablefoottext{c}{\citet{2016ApJ...822...97H}} \tablefoottext{d}{\citet{2018A&A...616A...1G}} \tablefoottext{e}{\citet{2008MNRAS.390..567M}} \tablefoottext{f}{\citet{2012ApJ...748...93R}} \tablefoottext{g}{\citet{2012ApJ...758...56S}} \tablefoottext{h}{\citet{2018MNRAS.479.4836L}}
            \tablefoottext{i}{\citet{2017NatAs...1E.184S}}.}
    \end{table}

\subsection{Activity indicators}

    High-resolution spectroscopy of activity diagnostics has revealed to be a powerful tool to improve our understanding of stellar chromospheres; optically thick photospheric lines with broad absorption wings have core emission features that are strictly linked to the chromosphere's thermal structure. High-resolution spectra are required to resolve these emission features and to characterise their complex profiles that often consist of emission peaks with a self-reversed dip at line centre.
    In particular, we analysed the fluxes and profiles of the \ion{H}{i} Balmer series, \ion{He}{i}, \ion{Na}{i}, and \ion{Ca}{ii} H\&K. 
    
    H$\alpha$ and \ion{Ca}{ii} K are two of the strongest optical emission lines in active M dwarf chromospheres. Across the M spectral class there is a range of emission strength in \ion{Ca}{ii} K, and a wide variety of both absorption and emission in H$\alpha$. The 
    H$\alpha$ core appears to trace hotter regions of the chromosphere ($\ge$ 7000 K), while \ion{Ca}{ii} K is formed in the cooler regions between the temperature minimum and $\sim$ 6000 K \citep{1982ApJ...258..740G, 1985ApJ...294..626C,  2009AJ....137.3297W}. Thus,   H$\alpha$ and \ion{Ca}{ii} K together offer complementary information on chromospheric structure. 
    
    The \ion{Ca}{ii} H (3968.47 \AA) and K (3933.66 \AA) lines are very useful diagnostics of the solar chromosphere. 
    The emission cores of the H\&K lines are weak for very quiet regions on the Sun, but can exceed the local continuum in brightness for active stars, particularly for active M dwarfs that have a weak continuum.
    For FGK stars, the H\&K lines show emission cores inside very broad absorption wings because \ion{Ca}{ii} is the primary ionisation stage in the photospheres and lower chromospheres of these warm stars. For M stars, \ion{Ca}{i} is the dominant ionisation stage in the photosphere and lower chromosphere, and as a consequence the H\&K lines for these stars do not have broad absorption wings \citep{2017ARA&A..55..159L}.
    
    Observations of the solar surface indicate that the inhomogeneities on the surface may be due to contributions from different regions and phenomena;  \ion{Ca}{ii} K core emission corresponds spatially to regions of concentrated magnetic field, such as  active plage regions and bright network grains, while H$\alpha$ chromospheric emission and absorption can be produced in filaments protruding from active regions, in spots across the network of the quiet Sun, and in enhanced emission from bright points during flares \citep{2008ApJ...680.1542H, 2006ASPC..354..276R, 2007ASPC..368...27R}. Consequently, examining the relationship between the \ion{Ca}{ii} and Balmer lines can   throw light on the nature of magnetic structures. 
    
    \citet{2017A&A...598A..28S}, extending a previous study by \citet{2010A&A...520A..79M}, analysed the short-term chromospheric variability and the flux excess emitted in the \ion{Ca}{ii} H\&K and H$\alpha$ lines of a sample of 71 early-type M dwarfs with different levels of activity (inactive and moderately active stars). They show that the \ion{Ca}{ii} H\&K flux excesses are strongly linearly correlated. When comparing the \ion{Ca}{ii} H\&K with the H$\alpha$ chromospheric line flux they found significantly more scatter, mostly for the most active stars. The same sample of inactive and moderately active stars was analysed by \citet{2017A&A...598A..27M}, who  focused on average trends.
    
    The sodium resonance doublet is an important photospheric and chromospheric diagnostic. The typical profile of \ion{Na}{i} doublet shows extended wings and narrow cores. Active dwarfs with H$\alpha$ in emission have been shown to exhibit a distinctive core emission of probable chromospheric origin \citep[e.g.][]{1978ApJ...226..144G,1981ApJS...46..159W,1991A&AS...90..437P}. \citet{1989A&A...209..279P} was the first to detect the important chromospheric contribution of the \ion{Na}{i} D$_{1,2}$ lines in the core for active M dwarfs.
    A complete study of the formation of the \ion{Na}{i} D$_{1,2}$ lines proposed by \citet{1997A&A...322..266A} confirmed that these lines are promising diagnostics of the lower-middle chromosphere. 
    \citet{2009MNRAS.400..238H} also shows that the main chromospheric contribution of these indicators arises in a narrow line core, but they also note some differences in the inner wings, suggesting that magnetic activity could also affect the upper photosphere. 
    
    The \ion{He}{i} D$_3$ (5875.62 \ \AA) is also an interesting diagnostic because it is formed in the lower transition region and it is mostly detected in very active stars.  All these chromospheric lines are used in planet search programmes to identify stellar activity, and they are all correlated to some extent with the RV jitter \citep[e.g.][]{2012A&A...541A...9G}.
    
    In this paper we present a study of all these chromospheric lines and their variability due to magnetic activity, focusing our attention on a specific M dwarf, well known for its high level of magnetic activity.
    
\section{Observations}\label{sec:obs}
   
    The high-resolution spectra of AD Leo analysed in this work were obtained with two different instruments. 
    We analysed 33 high-resolution spectra of AD Leo collected with HARPS \citep{2003Msngr.114...20M}, the fibre echelle spectrograph installed on the 3.6 m  European Southern Observatory (ESO) telescope in the La Silla Observatory, Chile, obtained from January to May 2006. 
    In addition we considered 63 HARPS-N \citep{2012SPIE.8446E..1VC} spectra collected in the context of the Global Architecture of Planetary System (GAPS) programme \citep{2013A&A...554A..28C}   \footnote{AD Leo was originally part of the search of planets around young stars of the GAPS 2 programme since a candidate planet around was proposed by \citet{2018AJ....155..192T} and then discarded by \citet{2020A&A...638A...5C}}. 
    HARPS-N observations were performed in two different observing seasons: from April to June  2018 and from November 2018 to January 2019. All the data used in this work are listed in Table \ref{tab:ObsData}.
        
    The two instruments have very similar performance with a resolving power of R $\sim$ 120000 (HARPS) and R $\sim$ 115000 (HARPS-N) and a spectral coverage of 378-691 nm and 383-693 nm, respectively. The spectra are provided already reduced using ESO/HARPS-N standard calibration pipelines.
    
\section{Analysis of the observations}\label{sec:analysis}
    We identified a number of lines sensitive to activity, listed in Table \ref{tab:rangeline}. A strong emission is detected, even during the quiescent state of the star, for the H$\alpha$, H$\beta$, \ion{Ca}{ii} H\&K lines; an intermediate emission above the continuum is observed for the He lines (\ion{He}{i} D$_3$, \ion{He}{i} 4026 \AA \ and \ion{He}{i} 4471 \AA ); and the \ion{Na}{i} doublet  (D$_1$ \& D$_2$) shows  emission in the core of the line profile.  
    
    These lines result from different excitation potentials, so their formation requires different physical conditions that occur in different parts of the active atmosphere of AD Leo. As a result, changes in equivalent width and/or in line profile of these lines can be explained by a direct or indirect impact of the magnetic activity on the whole stellar atmosphere and on its time variability.
    
    As a measure of the chromospheric activity strength, we measured excess of fluxes, as described in the next sections. 
    To measure the emission caused by activity, we chose wavelength integration ranges that are  sufficiently broad for the broadest emission even in case of a strong flaring event. These ranges were set after a visual inspection of the spectra and are reported in Table \ref{tab:rangeline} for each line we considered.
    
    In addition, other lines known as good indicators of chromospheric activity, such as the \ion{Mg}{i} b$_1$, b$_2$, b$_4$ lines and \ion{Fe}{i} at 5270 \AA, were analysed, showing the same behaviour as the other lines studied in this work, even though  their emission above the continuum is less intense than for the other lines, and for this reason they are not reported. 
    
    \begin{table}[h]\scriptsize
        \begin{center}
                \caption{Rest wavelength and integration ranges for the selected lines. Blue and red integration ranges were chosen to fit the continuum.}
                \label{tab:rangeline}
                \begin{tabular}{lcccc}
            \toprule[0.05cm]
            \toprule
                        Line    &       $\lambda$       &       Blue integration  & W &  Red integration         \\
                                &        (\AA)  &       ranges (\AA)    & (\AA)& ranges (\AA)     \\
            \midrule
                        \ion{Ca}{ii} K          &       3933.66 &       3932.20 - 3933.20       &       3933.20 - 3934.50       &       3934.50 - 3935.00       \\
\ion{Ca}{ii} H          &       3968.47 &       3967.70 - 3968.00       &       3968.00 - 3969.10       &       3969.10 - 3969.30       \\
\ion{He}{i} 4026                &       4026.19 &       4025.40 - 4026.10       &       4026.10 - 4026.70       &       4026.70 - 4027.00       \\
\ion{He}{i} 4471                &       4471.48 &       4470.00 - 4471.40       &       4471.40 - 4471.85       &       4471.85 - 4473.00       \\
H$\beta$                &       4861.35 &       4858.70 - 4859.60       &       4859.60 - 4864.00       &       4864.00 - 4864.20       \\
\ion{He}{i} 5876                &       5875.62 &       5875.30 - 5875.42       &       5875.28 - 5876.80       &       5876.90 - 5877.00       \\
\ion{Na}{i} D$_2$               &       5889.95 &       5889.50 - 5889.80       &       5889.80 - 5890.70       &       5890.70 - 5891.00       \\
\ion{Na}{i} D$_1$               &       5895.92 &       5895.70 - 5895.80       &       5895.90 - 5896.50       &       5896.60 - 5896.70       \\
H$\alpha$               &       6562.79 &       6553.00 - 6555.00       &       6555.00 - 6570.00       &       6570.00 - 6572.00       \\
            \bottomrule[0.05cm]
        \end{tabular}
    \end{center}
    \end{table}

    \begin{figure*}[ht!]
    \resizebox{\hsize}{!}
            {\includegraphics[width=\hsize]{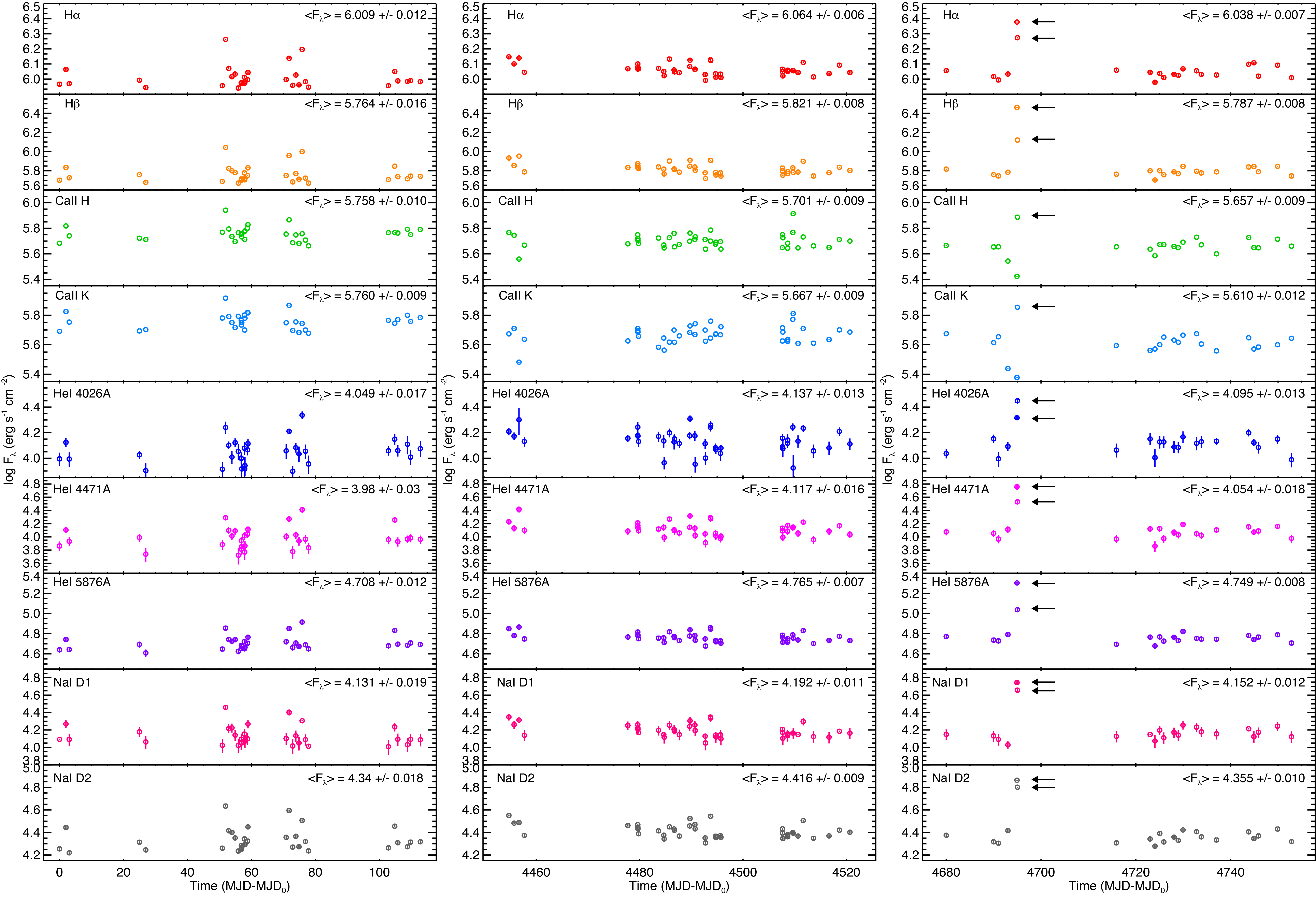}}
            \caption{Line flux vs time (MJD$_0$ is the start time of observations in 2006). Data obtained in 2006 are shown in the left panel. Data obtained in 2018 are shown in the middle and right panels. <F$_\lambda$> is the average of the logarithmic flux of each activity indicator for each season (for the second season of 2018 these values   exclude the  flare event points). Black arrows mark the points relative to the flare event. The error bars are shown in the plots, but for most of the points are too small to be visible.}
         \label{fig:Timeseries}
    \end{figure*}
    \subsection{Flux rescaling}\label{sec:flux-rescaling}
    The HARPS and HARPS-N spectra are not calibrated in flux; therefore, they have arbitrary units.  
    The spectra provided by the data reduction system (DRS) show night-to-night variations in the continuum level at different wavelengths, due to atmospheric differential absorption and instrumental effects. To correct them, and to scale the observed spectra to the same flux reference, in order to be able to compare the intensity of the analysed lines, we compare them with synthetic spectra from the BT-Settl spectral library provided by \citet{2011ASPC..448...91A}\footnote{We adopt the CIFIST2011 models (\url{https://phoenix.ens-lyon.fr/Grids/BT-Settl/CIFIST2011bc/SPECTRA/})} with $T_{\mathrm{eff}}$, log(\textit{g}), and [Fe/H] corresponding to stellar parameters (see Table \ref{table:target}) in analogy to the procedure adopted to compute the excess fluxes provided by \citet{2017A&A...598A..28S}. 
    Both the observed and the model spectra were degraded to low resolution, convolving them with a Gaussian kernel with $\sigma = 80$ \ \AA, in order to avoid discrepancies between the observed and the model lines profiles.
    Finally, the observed-to-model flux ratio was used to rescale the observed high-resolution spectra.

    The flux calibration procedure may be less accurate in the case of strong emission lines, sensitive to the magnetic activity, because the model does not take into account the chromospheric emission; therefore, to obtain a more precise calibration in those areas, they are removed during this procedure.
    
    We used the flux calibrated spectra to calculate the flux for each line according to Eq \eqref{eq:Fline} with the same integration ranges listed in Table \ref{tab:rangeline}. This value provides a measure similar to the equivalent width (EW), but less influenced by continuum flux estimation. This is important for lines located in spectral regions where the continuum is very low, and hence its relative uncertainty is very high. The flux line is computed as

    \begin{equation}
        \label{eq:Fline}
        F_{line} = \sum_{i=1}^{i=n} F_{i}d\lambda - \dfrac{(F_{c,b}+F_{c,r})}{2}W
    ,\end{equation}
    where d$\lambda$ is the width of the wavelength bin; $F_{i}$ is the observed flux in the bin $i$ of the line; $n$ is the number of bins within the line region, defined as $W/d\lambda$; F$_{c,b}$ and F$_{c,r}$ are the flux values measured at the extremes of the integration range on the blue and red side of the line, respectively; and $W$ is the wavelength range used for the integration, corresponding to the full line width (see Table \ref{tab:rangeline}).
    
    Several tests were done to find the most accurate method for determining the continuum flux F$_c$. We chose to fit the continuum (in the blue and red integration ranges defined in Table \ref{tab:rangeline}) with a linear function. This method shows that the continuum flux is, with good approximation, constant over the considered range in most of the analysed spectra; however, some spectra show a slope and the linear fit allows us to take it into account. The error of the continuum flux, $\delta$F$_{c,i}$, was   estimated by applying the standard error propagation theory on the uncertainties of the fit parameters.
    
    There are no obvious estimates for the statistical error of the observed flux, $\delta$F$_{i}$. The spectrum is affected by the presence of numerous minor lines that are not variable in time, and that characterise every part of the spectrum. Since these lines are too numerous to   be isolated, and since they can affect the spectrum in the continuum and  in the profile of the line, we can assume that the $\delta F_{i}$ is the standard deviation with respect to the continuum flux calculated in the N points outside the line (Eq.\ref{eq:errflusso}):
    
     \begin{equation}\label{eq:errflusso}
        \delta F_{i} = \sqrt{\dfrac{\sum_i \big(F_{i} - F_{c,i}\big)^2}{N-1}}
    .\end{equation}
    
    The $F_{line}$ uncertainty was estimated using Eq. \ref{eq:errflussoline}, assuming $d\lambda = 0.01$ \AA:
    
    \begin{equation}\label{eq:errflussoline}
        \delta F_{line} = \sqrt{\sum_i\bigg(\delta F_{i} \ d\lambda\bigg)^2 + \bigg(\delta F_{c,b}^2+\delta F_{c,r}^2\bigg)\bigg(\dfrac{W}{2}\bigg)^2+\delta F_{range}^2}~.
    \end{equation}Here F$_{range}$ takes into account the possible effects due to the selection of the ranges used to estimate the continuum ($\delta W$). This value was calculated as the half difference between the maximum and minimum values of the continuum flux obtained with three different ranges for the continuum measurements. 
    
    
    \subsection{Time series and line flux  variability}
    Figure \ref{fig:Timeseries} shows the temporal variations of the analysed activity indicators. In particular, Fig. \ref{fig:Timeseries} shows the variability of the integrated line flux of the analysed lines with time. The left panel shows HARPS data obtained in 2006;  the middle and right panels show two different observing seasons of HARPS-N dataset performed in 2018. 

    Several conclusions can be drawn from this figure. First, we can confirm that the flux on the stellar surface for the analysed lines is variable on both short (hours, days) and long (months, years) timescales during the entire observed time. Second, during the second season of 2018 (right panel) a flare is observed: two points, corresponding to two observations obtained two hours apart,  highlight this phenomenon and allow us to follow its evolution. A more detailed analysis of the flare is described in Sect. \ref{section:flare}.  Three other possible flare events are detected during 2006 (left panel).    
    Moreover, observing the time series of the analysed activity indicators, we can assert that AD Leo was more active in 2018 than in 2006. Unexpectedly, despite the lower activity level, the time series of \ion{Ca}{ii} H\&K show a higher  flux in 2006 (see the average of the logarithmic flux <F$_\lambda$> in Fig. \ref{fig:Timeseries} and the histograms in Fig. \ref{fig:Corr20062018}).
    
    \section{Flux-flux relationship}\label{sec:flux-flux}
\begin{figure*}
    \centering
            \includegraphics[scale=0.35]{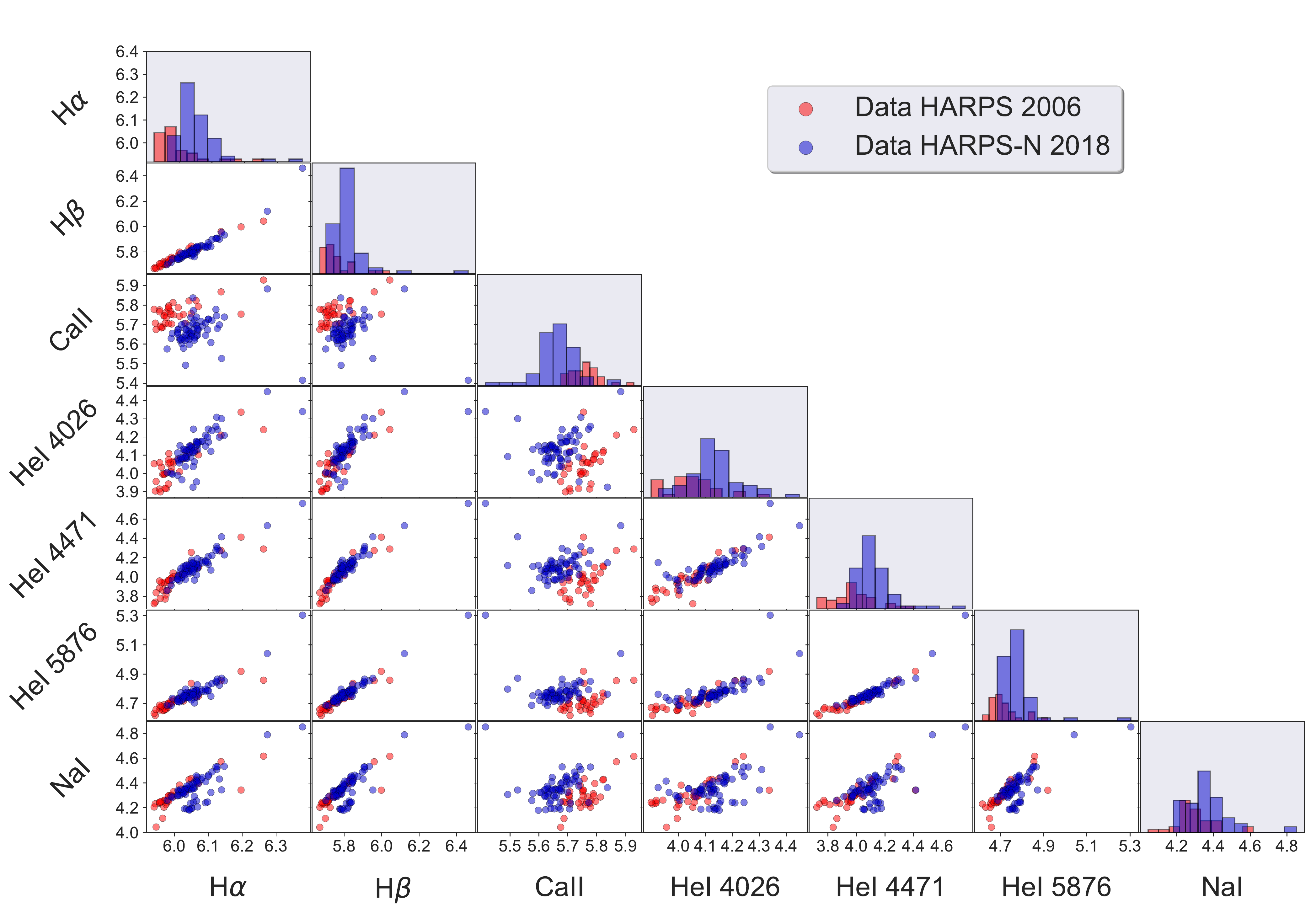}
            \caption{Correlation plot of flux (logarithmic scale) between different activity indicators. The diagonal panels show the histogram of flux of each indicator. It can be seen that most of the indicators show a significant correlation in both the datasets. Correlations with \ion{Ca}{ii} are less significant and more scattered. This figure also shows less activity of the star in 2006 (red points) than in 2018 (blue points). However, the flux of \ion{Ca}{ii} is higher in 2006 than in 2018. This result can be interpreted as a major surface coverage of plages and filaments during the observations in 2006.}
         \label{fig:Corr20062018}
    \end{figure*}

    In the following we analyse the relationships between the chromospheric fluxes of different activity indicators.
    We inspect the presence of a correlation based on Spearman’s rank-order correlation coefficient ($\rho$).
    Figure \ref{fig:Corr20062018} shows the correlations between the fluxes obtained from observations in 2018 (dark blue points) and those in 2006 (red points), and the results of the statistical tests separated for the two seasons are provided in Table \ref{tab:correlation}. 

    \begin{table}\scriptsize
        \begin{center}
                \caption{Statistical analysis of chromospheric activity indicators fluxes. In the third and fourth columns we reported the Spearman coefficient $\rho$ and the value of the probability of a null hypothesis for the  dataset of 2006. In the last two columns we reported the same values for the dataset of 2018. Weak correlations are reported in brackets, no correlations are shown in boldface.}
                \label{tab:correlation}
                \begin{tabular}{llcccc}
            \toprule[0.05cm]
            \toprule
                            X-index     &       Y-index & $\rho_{2006}$\tablefootmark{a}  & P$_{\rho_{2006}}$\tablefootmark{b}&  $\rho_{2018}$ & P$_{\rho_{2018}}$ \\
            \midrule
                H$\alpha$       &       H$\beta$        &       0.965   &       0.000006        \%      &       0.941   &       0       \%      \\
                H$\alpha$       &       \ion{Ca}{ii} H  &       0.518   &       0.34    \%      &       0.490   &       0.01    \%      \\
                H$\alpha$       &       \ion{Ca}{ii} K  &       (0.438) &       (1.33   \%)     &       \textbf{0.233}  &       \textbf{6.4     \%}     \\
                H$\alpha$       &       \ion{Ca}{ii}    &       0.494   &       0.5     \%      &       0.409   &       0.12    \%      \\
                H$\alpha$       &       \ion{He}{i} 4026        &       0.816   &       0.0004  \%      &       0.769   &       0       \%      \\
                H$\alpha$       &       \ion{He}{i} 4471        &       0.920   &       0.000018        \%      &       0.831   &       0       \%      \\
                H$\alpha$       &       \ion{He}{i} 5876        &       0.953   &       0.000006        \%      &       0.802   &       0       \%      \\
                H$\alpha$       &       \ion{Na}{i}     &       0.932   &       0.000012        \%      &       0.585   &       0.0003  \%      \\
                H$\beta$        &       \ion{Ca}{ii} H  &       0.532   &       0.26    \%      &       0.459   &       0.027   \%      \\
                H$\beta$        &       \ion{Ca}{ii} K  &       (0.433) &       (1.43   \%)     &       \textbf{0.212}  &       \textbf{9.26    \%}     \\
                H$\beta$        &       \ion{Ca}{ii}    &       0.489   &       0.6     \%      &       0.390   &       0.2     \%      \\
                H$\beta$        &       \ion{He}{i} 4026        &       0.821   &       0.0003  \%      &       0.831   &       0       \%      \\
                H$\beta$        &       \ion{He}{i} 4471        &       0.952   &       0.000006        \%      &       0.886   &       0       \%      \\
                H$\beta$        &       \ion{He}{i} 5876        &       0.948   &       0.000006        \%      &       0.901   &       0       \%      \\
                H$\beta$        &       \ion{Na}{i}     &       0.933   &       0.000012        \%      &       0.601   &       0.00018 \%      \\
                \ion{Ca}{ii} H  &       \ion{Ca}{ii} K  &       0.928   &       0.000018        \%      &       0.771   &       0       \%      \\
                \ion{Ca}{ii} H  &       \ion{He}{i} 4026        &       0.597   &       0.07    \%      &       0.338   &       0.73    \%      \\
                \ion{Ca}{ii} H  &       \ion{He}{i} 4471        &       0.543   &       0.21    \%      &       0.379   &       0.26    \%      \\
                \ion{Ca}{ii} H  &       \ion{He}{i} 5876        &       0.517   &       0.3     \%      &       0.408   &       0.12    \%      \\
                \ion{Ca}{ii} H  &       \ion{Na}{i}     &       0.517   &       0.43    \%      &       0.520   &       0.004   \%      \\
                \ion{Ca}{ii} K  &       \ion{He}{i} 4026        &       0.511   &       0.38    \%      &       \textbf{0.123}  &       \textbf{33      \%}     \\
                \ion{Ca}{ii} K  &       \ion{He}{i} 4471        &       (0.422) &       (1.69   \%)     &       \textbf{0.187}  &       \textbf{14      \%}     \\
                \ion{Ca}{ii} K  &       \ion{He}{i} 5876        &       (0.409) &       (2.1    \%)     &       \textbf{0.191}  &       \textbf{13      \%}     \\
                \ion{Ca}{ii} K  &       \ion{Na}{i}     &       (0.404) &       (2.2    \%)     &       0.327   &       0.94    \%      \\
                \ion{Ca}{ii}    &       \ion{He}{i} 4026        &       0.591   &       0.08    \%      &       \textbf{0.268}  &       \textbf{3.37    \%}     \\
                \ion{Ca}{ii}    &       \ion{He}{i} 4471        &       0.494   &       0.52    \%      &       0.336   &       0.76    \%      \\
                \ion{Ca}{ii}    &       \ion{He}{i} 5876        &       0.479   &       0.69    \%      &       0.344   &       0.63    \%      \\
                \ion{Ca}{ii}    &       \ion{Na}{i}     &       0.478   &       0.69    \%      &       0.479   &       0.014   \%      \\
                \ion{He}{i} 4026        &       \ion{He}{i} 4471        &       0.86    &       0.00009 \%      &       0.844   &       0       \%      \\
                \ion{He}{i} 4026        &       \ion{He}{i} 5876        &       0.842   &       0.00019 \%      &       0.808   &       0       \%      \\
                \ion{He}{i} 4026        &       \ion{Na}{i}     &       0.777   &       0.0011  \%      &       0.397   &       0.16    \%      \\
                \ion{He}{i} 4471        &       \ion{He}{i} 5876        &       0.943   &       0.000012        \%      &       0.919   &       0       \%      \\
                \ion{He}{i} 4471        &       \ion{Na}{i}     &       0.887   &       0.00005 \%      &       0.478   &       0.015   \%      \\
                \ion{He}{i} 5876        &       \ion{Na}{i}     &       0.947   &       0.000006        \%      &       0.536   &       0.002   \%      \\
                \ion{Na}{i} D1  &       \ion{Na}{i} D2  &       0.867   &       0.00001 \%      &       0.893   &       0       \%      \\
            \bottomrule[0.05cm]
        \end{tabular}
    \end{center}
    \tablefoot{\tablefoottext{a}{Rank correlation for two populations} \tablefoottext{b}{P-value denotes the two-sided significance of its deviation from 0 by random chance, i.e. a small value indicates significant correlation}.}
    \end{table}
    
    It can be seen that most of the indicators show a significant correlation (P $<1\%$) in both datasets. The \ion{Ca}{ii} K line has a peculiar behaviour, with a weak correlation ($1 \% <$ P $<3\%$) in the 2006 season and no correlation (P $>3\%$) in 2018 with the other indicators. The correlation between most of the analysed lines implies that they have a similar origin and are likely formed from the same material or from the same region of the star's atmosphere.
    
    We evaluated the same correlations excluding the points relative to the flare to verify their impact. We found that they do not influence the correlations among the indices.

    Finally, we verified that the correlation among the activity indices for the whole dataset is maintained when we join data obtained 12 years apart, with the only exception of the \ion{Ca}{ii} H\&K index, for which there is no correlation with the other indices.
    
    In addition, we estimated the Balmer decrements (H$\alpha$, H$\beta$),  which are indicators of the physical conditions of the emitting regions \citep[e.g.][]{1979ApJ...230..581L,1991PhDT.........9C}. \citet{2017A&A...598A..27M} showed the Balmer decrement as a function of the effective temperature and overplotted the typical values of solar plages. Our result ($\sim 1.76$) is compatible with values of solar plages, suggesting that AD Leo is dominated by them.

    \subsection{\ion{Ca}{ii} H\&K versus H$\mathrm{\alpha}$}
    The comparison between H$\alpha$ (or H$\beta$) and \ion{Ca}{ii} (\ion{Ca}{ii} H\&K) fluxes shows that the correlation between these two indicators is less significant and more scattered than the correlations between the other lines. 
    
    This result is consistent with the hypothesis that the phenomena that produce the two lines are actually connected, but the materials that generate them are in different regions of the atmosphere. Moreover, we also tested the correlations between the indicators excluding the measurements taken during the flare. These further tests return a value only slightly more significant than the previous one. 
    
\begin{figure}[h!]
    \centering
    \includegraphics[width=0.95\hsize]{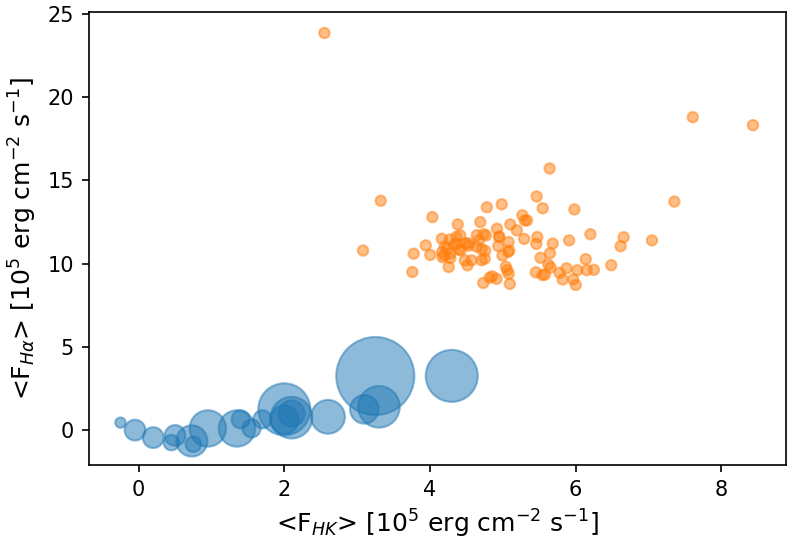}
    \caption{Plot of F$_{HK}$ vs F$_{H\alpha}$. The blue bubbles map the region populated by the stars analysed by \cite{2017A&A...598A..28S}, the orange points are the flux values obtained for AD Leo in this study.}
    \label{fig:rico_scandariato}%
    \end{figure}

    The result obtained from the test is consistent with those presented by  \cite{2017A&A...598A..28S}, who show that H$\alpha$ and \ion{Ca}{ii} H\&K are correlated and that the correlation is more scattered for the most active stars. Specifically, in Fig. \ref{fig:rico_scandariato}, the blue bubbles represent an envelope of the results obtained by \cite{2017A&A...598A..28S}, while the orange points are the values obtained for AD Leo in this study.
    
    Furthermore, although all the other activity indicators are more intense in 2018, the flux of \ion{Ca}{ii} H\&K is higher in 2006 than in 2018. By considering the model of \citet{2009A&A...501.1103M}, which   affirms that \ion{Ca}{ii} core emission is connected to the active plage regions and bright network grains, while the H$\alpha$ line is produced from all the inhomogeneities present on the stellar surface, our result can be interpreted with a major surface coverage of plages and filaments during the observations on 2006.
    Even though the Balmer decrement suggests that AD Leo is dominated by plages, this ratio does not allow us to distinguish between the two observing seasons.

    \section{Flare analysis}
    \label{section:flare}
    Solar and stellar flares are  observable evidence of magnetic energy released on  short timescales. The magnetic reconnection plays a key role in the reconfiguration of the magnetic field lines and the conversion of magnetic energy into kinetic and thermal energies of plasma \citep{1996ApJ...459..330F, 2000mare.book.....P}. The impulsive X-ray and UV emission associated with stellar flares can affect the stellar atmosphere.
    
    The most extreme solar flare that hit Earth was recorded in 1859 \citep{1859MNRAS..20...13C,1859MNRAS..20...15H}. It released a flare energy of $10^{32}$ erg. 
    Stellar flares are expected to be generated by the same mechanism of solar flares with a wider range of energy radiation and timescale \citep[e.g.][]{2010ARA&A..48..241B,2018MNRAS.475.2842D}. Over short timescales of minutes to a few hours, they emit energy ranging from $10^{23}$ erg (called nanoflares) \citep[e.g.][]{2000ApJ...529..554P} to $10^{33}-10^{38}$ erg (called superflares) \citep[e.g.][]{2013ApJS..209....5S}.
    
    From the standard solar flare model, flares are formed by accelerated non-thermal electrons that propagate downward and heat the chromosphere. As a consequence, the heated chromospheric material moves upward (evaporation), filling the  coronal loop above. This material then cools down radiating away its excess energy, and finally moves downward (condensation), going back to  the lower layers of the stellar atmosphere \citep{1998ApJ...494L.113Y}. Because of the high temperatures and large motions of the flaring material, chromospheric emission lines during flares appear much broader than in the quiescent state of star.
    
    In the right panels of Fig. \ref{fig:Timeseries} we indicate with black arrows two consecutive points obtained during the second observing season of 2018, where the flux of all activity indicators is significantly higher than the quiescent state of the star. Therefore, it is reasonable to assume the presence of a flare. 
    Since the two spectra were obtained two hours apart, we have the possibility to follow roughly the temporal evolution of the flare. We can suppose that the first observation during the flare is relative to the maximum phase of the flare, while the second point, with lower value of flux than the first one, was obtained during the decaying phase of the flare. 
    
    The observed profiles of some selected spectral lines sensitive to the stellar activity are broadened during the flare. This can be due to the motion of material inside the magnetic loop. 
    
    We considered a number of lines where the broadening is more evident (H$\alpha$, H$\beta$, \ion{He}{i} 4471 \AA, \ion{He}{i} 5876 \AA) and  we fitted each profile with two Gaussian components \citep[see][]{2006A&A...452..987C,2018A&A...615A..14F}. The Balmer lines show a self-reversal absorption in the core, but this behaviour was not taken into account because it does not have a significant contribution on the following analysis of the flare. The fit with two components results in a reasonably good description of the line profile even in the most asymmetric cases. In general, the Balmer lines display two distinct phases, called the impulsive and the gradual phases, with broader profiles during the impulsive phase and narrower profiles during the gradual phase. We do not consider the \ion{Ca}{ii}
    H\&K even if they are strong emission lines because they are not significatively influenced by the flare and they do not show broadening. Because the flare event is supposed to be generated in different regions with respect to the plages, the fact that \ion{Ca}{ii} lines are not broadened is consistent with the hypothesis that this indicator is influenced by the presence of plages and that AD Leo is dominated by them. 
    The results of the fit (the redshift and the sigma) for the narrow and the broad components are provided in Table \ref{tab:valorishift}. 
    
    \begin{table}[h]\footnotesize
        \begin{center}
                \caption{Fitted value of redshifts $\delta v$ and sigma $\sigma(\delta v)$ of the narrow and broad components for ID 79 and ID 80 spectra taken during the flare. The errors resulting from the fit are $\le 0.1\%$.}
                \label{tab:valorishift}
                \begin{tabular}{lcccc}
            \toprule[0.05cm]
                        \toprule
                        \multicolumn{5}{c}{ID obs 79} \\
                        \midrule
                        \multirow{3}*{Line} & \multicolumn{2}{c}{Narrow} & \multicolumn{2}{c}{Broad} \\
                        \cmidrule(lr){2-5}
                        & $\delta v$ & $\sigma(\delta v)$ & $\delta v$ & $\sigma(\delta v)$ \\
                        & (km s$^{-1}$) &  (km s$^{-1}$) & (km s$^{-1}$) &  (km s$^{-1}$)  \\
                        \midrule
                    H$\alpha$   &       0.55    &       31.10   &       1.77    &       155.38  \\
            H$\beta$    &       1.50    &       26.51   &       4.31    &       129.03  \\
            \ion{He}{i} 4471    &       0.68    &       7.93    &       15.77   &       16.73   \\
            \ion{He}{i} 5876    &       2.06    &       8.21    &       10.13   &       15.49   \\
                    \midrule[0.04cm]
                    \multicolumn{5}{c}{ID obs 80} \\
                        \midrule
                        \multirow{3}*{Line} & \multicolumn{2}{c}{Narrow} & \multicolumn{2}{c}{Broad} \\
                        \cmidrule(lr){2-5}
                        & $\delta v$ & $\sigma(\delta v)$ & $\delta v$ & $\sigma(\delta v)$ \\
                        & (km s$^{-1}$) &  (km s$^{-1}$) & (km s$^{-1}$) &  (km s$^{-1}$)  \\
                        \midrule
                    H$\alpha$   &       0.68    &       28.82   &       29.52   &       170.99  \\
            H$\beta$    &       1.48    &       22.34   &       34.95   &       86.30   \\
            \ion{He}{i} 4471    &       0.96    &       6.38    &       21.49   &       11.00   \\
            \ion{He}{i} 5876    &       0.28    &       8.72    &       15.78   &       9.35    \\
                    \bottomrule[0.05cm]
    \end{tabular}
    \end{center}
    \end{table}
    
    \begin{figure*}[h!]
    \centering
    \includegraphics[width=0.8\hsize]{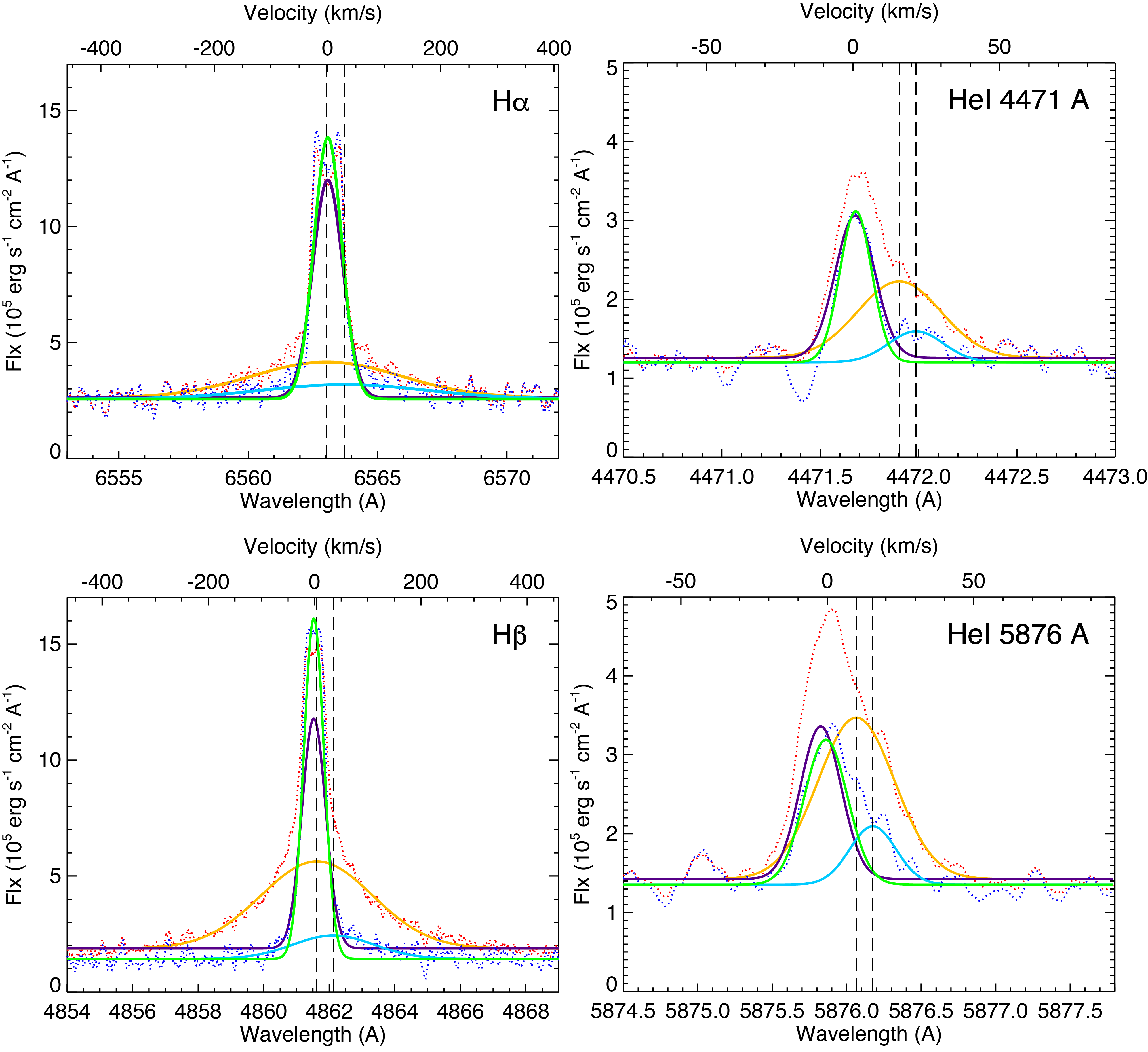}
    \caption{Spectrum ID 79   for the flare's maximum phase (red dotted line) and spectrum ID 80 for the decay phase (blue dotted line). Gaussian fit with   broad and narrow components, respectively  in orange and purple for ID 79 and in green and light blue for ID 80. The black dashed line shows the centre of each broad component.}
    \label{fig:fit}%
    \end{figure*}
    
    Figure \ref{fig:fit} shows the fits that we  made on spectra obtained during the flare. The red dotted line corresponds to the spectrum obtained during the maximum phase of the flare (ID 79), the blue dotted line to the spectrum obtained during the decaying phase (ID 80) of the flare.    The orange and light blue Gaussians represent the broader components for the observation ID 79 and ID 80, respectively, while the purple (ID 79) and green (ID 80) are the narrow components obtained from the fit.
    
    The spectra in Fig. \ref{fig:fit} show that the broadening of Balmer lines is larger than that of the helium lines. 
    The broad components of the  Balmer and helium lines are more redshifted than the narrow components. \citet{1988A&A...193..229D} observed a similar effect during flare on YZ CMi and suggest the presence of material inside the loop corresponding to different flare kernels that brighten successively one after another. Each downflow would produce a redshifted contribution to the Balmer lines. 
    
    Moreover, Fig. \ref{fig:fit} shows symmetric broadening during the decay phase (light blue component), for H$\alpha$ and H$\beta$, with $\sigma$ of the order of hundreds of km/s.
    This symmetric broadening can be interpreted with the presence of material inside the magnetic loop that undergoes blueshift and redshift simultaneously. The exposure time (900 seconds) of the observations obtained with HARPS-N, shorter than the evolution time of the flare, leads us to exclude the possibility that we are monitoring the same material before going uphill inside the loop and then downhill. 
    This result can be explained instead as the presence of turbulent motion that can be dominant with respect to the coherent motion of the material (uphill or downhill) \citep[see][]{1999ASPC..158..226M,2005A&A...436..677F}. 
    H$\alpha$ monitors the lower regions of the magnetic loop; in this region, due to the high density of the material, the turbulent motion can be dominant with respect to the coherent motion of the material, which  instead follows the magnetic field lines. 
    Globally, the lines are shifted due to the coherent motion, but the broadening due to the turbulence is much larger and dominates the shape of the line.

    On the contrary, Fig. \ref{fig:fit} (right panels) shows an asymmetrical broadening of helium lines with velocity of the order of tens of km s$^{-1}$. This asymmetric broadening might be present even in the Balmer lines, but it is clearly smaller than the symmetric broadening shown in H$\alpha$ and H$\beta$ and for these reasons it cannot be detected. We can suppose that the helium lines monitor an upper region of the loop higher than H$\alpha$. If in the lower chromospheric regions the kinetic energy density of the turbulent motion is probably comparable to the magnetic energy density, in the upper regions the magnetic energy density dominates the kinetic energy density making the motion of the plasma less turbulent and inducing it to move along the magnetic field lines. 
    This effect leads to a decrease in the line broadening and emphasises  the radial velocity shift.

    In addition, despite the low temporal resolution, we  identified a delay of a flare event for the \ion{Ca}{ii} H\&K and \ion{He}{i} at 4026 \AA \ with respect to Balmer lines. The moment at which a line reaches its maximum is related to the temperature that characterises the formation of the line, and therefore it is also related to the height at which the line is formed. Therefore, we can suppose that this delay, also observed by \citet{2006A&A...452..987C}, confirms that these lines monitor different regions of the stellar atmosphere with respect to the Balmer lines.
    
    We also tried  to estimate the luminosity and the energy released during the flare. According to our data, the line luminosity, estimated by analysing the lines during the flare, is significantly higher than the luminosity of the quiescent state of the star. The energy released ($\sim 10^{30}$ erg to $\sim 1.4 \times 10^{32}$ erg for the Balmer lines) is consistent with the presence of a particularly intense flare event, stronger than the flares detected by \citet{2006A&A...452..987C} who obtained an energy released value of the order of $10^{29}$ erg. In support of our results, we mention that \citet{2019arXiv191109922G}, observing AD Leo for 222 hours with the Echelle spectrograph of the 2 m telescope Alfred-Jensch-Teleskope in Tautenburg, detected 22 flares,  the largest of which emitted 2.9 $10^{31}$ erg in H$\alpha$ and 1.8 $10^{32}$ erg in H$\beta$.  \citet{2020arXiv200306163M}, analysing more than 2000 spectra of AD Leo collected with the same telescope in the context of the flare-search programme of the Th\"{u}ringer Landessternwarte, also detected numerous flares;  the largest one emitted 8.32 $10^{31}$ erg in H$\beta$ and 2.12 $10^{32}$ erg in H$\alpha$. Results from both studies are comparable to the energy released by our flare.
    A more detailed analysis of the flare is described in Appendices \ref{appendice_delay} and \ref{appendice_energy}.

    \section{Summary and conclusions}\label{sec:summary}
    
    In this paper we analysed the spectra of AD Leo using two datasets HARPS and HARPS-N spectra, obtained 12 years apart. We measured the line profiles and the intensities of the sensitive activity indicators, such as H$\alpha$, H$\beta$, \ion{Ca}{ii} H\&K, \ion{He}{i} at 4026 \AA, 4471 \AA, and 5876 \AA, and \ion{Na}{i} doublet. We derived the fluxes of these lines and evaluated the correlations between them.
    
    By analysing the time variability of the fluxes we found a higher level of activity during 2018 than 2006, except for the \ion{Ca}{ii} H\&K indicator that shows a higher flux on 2006. As suggested by \citet{2008ApJ...680.1542H} and \citet{2006ASPC..354..276R, 2007ASPC..368...27R}, the \ion{Ca}{ii} core emission originates from regions of concentrated magnetic field, such as active plages and bright grain networks. According to this, the longterm variability of \ion{Ca}{ii} suggests that the star had a larger coverage of plages during the observations of 2006 than in 2018.  
    Furthermore, the Balmer decrements (H$\alpha$/H$\beta$), calculated for the three observing seasons, are compatible with the typical values of solar plages showed by \citet{2017A&A...598A..27M}, confirming that the stellar surface is probably covered by a distribution of plages.  
    
    We searched for the correlation among the activity indicators measured in this work. All lines show a good correlation with each other, except for the \ion{Ca}{ii}, particularly the K line, indicating that the processes and regions of the formation of this line differ from other lines. 
    Many studies \citep[e.g.][]{2009AJ....137.3297W,2007A&A...469..309C} suggest that there is a correlation between H$\alpha$ and \ion{Ca}{ii} K flux obtained for a sample of different stars of different spectral types. However, \citet{2007A&A...469..309C} have declared that `when we investigate this relation for individual observations of a particular stars, the general trend is lost and each star shows a particular behaviour, ranging from tight correlations with different slopes, to anti-correlations, including cases where no correlations are found'. \citet{2009AJ....137.3297W} compared the equivalent width of H$\alpha$ to the \ion{Ca}{ii} K surface flux measured from a sample of M stars. They found a positive correlation between the measurements of these indicators when comparing different stars, with a wide range of scatter for the more active stars. Furthermore, they obtained multiple measurements of EW of Balmer lines and \ion{Ca}{ii} K in AD Leo and showed that for individual active stars these two lines are not necessarily correlated in time-resolved observations.
    Our flux values obtained for \ion{Ca}{ii} H\&K and H$\alpha$ follow the extrapolation of the trend shown in Fig. 10 of \citet{2017A&A...598A..27M}, confirming that the same trend continues at a high activity level.
    
    We also detected the presence of a flare during the second season of HARPS-N data. 
    \citet{2006A&A...452..987C} monitored AD Leo during four nights in 2001 and observed a large number of short and weak flares occurring very frequently. We  measured the EWs\footnote{The EWs were   measured with a procedure similar  to that  of fluxes (see Sect. \ref{sec:flux-rescaling}), except for the normalisation, and the results are provided in Appendix \ref{appendice:tabelle}.} of the analysed lines to compare our results to the published ones. The range of EW values that we obtained during the entire observed time identified as the `quiescent' state of the star is consistent with the variability of Balmer lines EWs obtained by \citet{2006A&A...452..987C}.
    Moreover, the surface fluxes of the Balmer lines at flare maximum (F$_{max}$) obtained by \citet{2006A&A...452..987C} are an order of magnitude lower than our results (see Table \ref{tab:luminositàenergiabrillamento}). This implies, also due to our low temporal resolution, that we are unable to resolve less intense flares and that what we call quiescent state is indeed the superposition of several weak flares. The flare that we  observed is a stronger and uncommon event. In this work we presented a detailed analysis of the profile of selected emission lines to study dynamic processes occurring during this phenomenon. In particular, we analysed the profiles of H$\alpha$, H$\beta$, and \ion{He}{i} at 4471 \AA \ and 5876 \AA \ from two spectra collected during the flare and obtained two hours apart, showing a significant broadening, while no evidence of broadening is present in the \ion{Ca}{ii} lines.
    We fitted the profiles combining a broad and a narrow Gaussian component, finding that the broader one is redshifted with a velocity of the order of tens of km s$^{-1}$. This redshift can be interpreted as the presence of material going downhill inside the magnetic loop, according to the solar flare model. Globally, the shape of these lines, especially for the Balmer lines, is symmetrically broadened with $\sigma$ of the order of hundreds of km s$^{-1}$. Since H$\alpha$ monitors the lower regions of the magnetic loop, we can suppose that in this region, because of the high density of the material, the turbulent motion can be dominant over the coherent motion of the material that follows the magnetic field lines. Consequently, we can suppose that  the Balmer lines are also redshifted due to the coherent motion of the material, but that this redshift is hidden by the broadening due to the turbulence that is much larger and dominates the shape of the lines.

    \balance{
    \bibliographystyle{aa}
    \bibliography{bibliography_AD_Leo}}

   \begin{appendix}
   \section{Delay of flare}\label{appendice_delay}
\begin{figure}[htp!]
    \centering
    \includegraphics[width=\hsize]{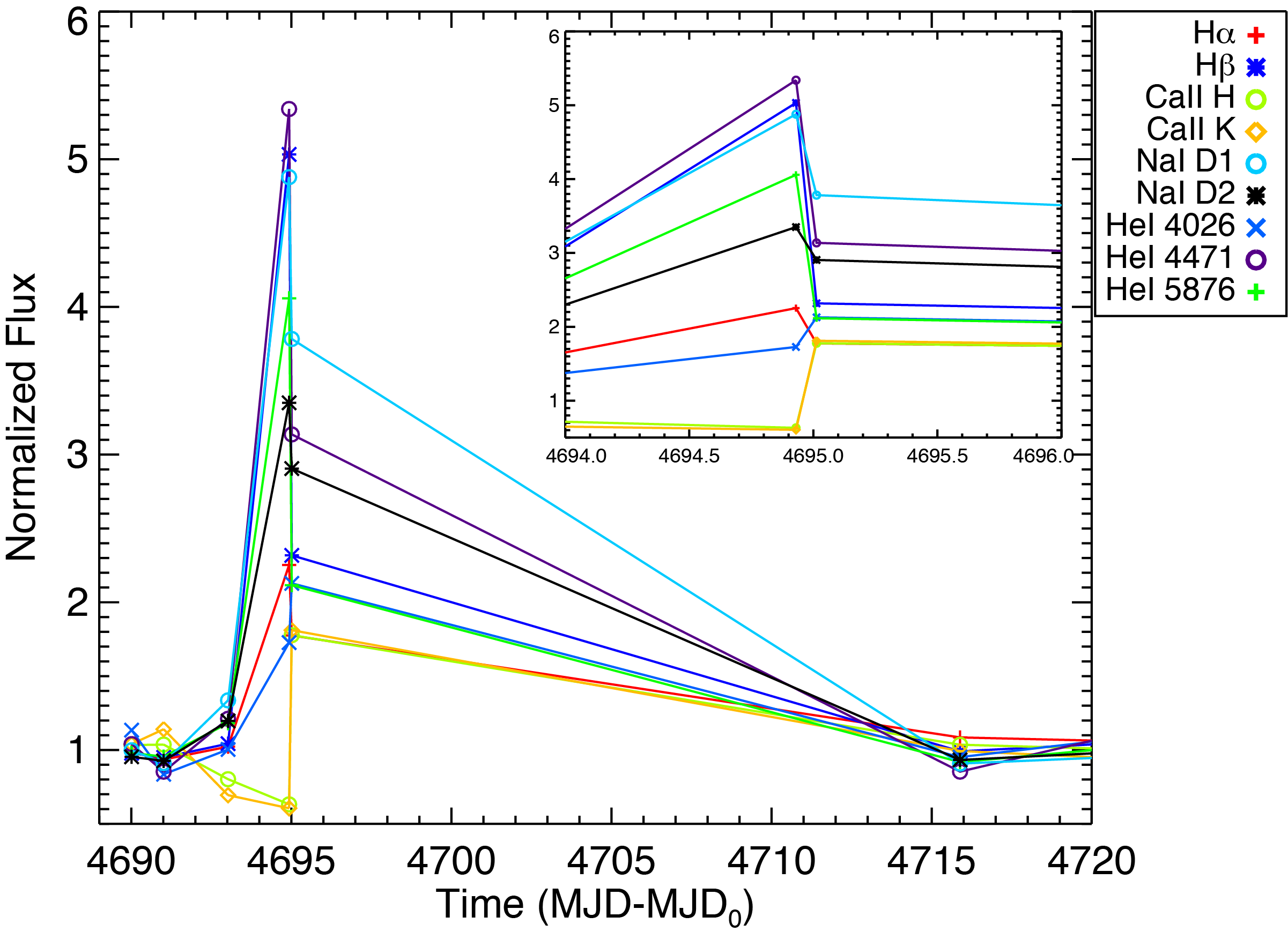}
    \caption{Time series of normalised flux of analysed activity indicators to evidence the flare. MJD$_0$ = 53758.244, time of the first observation obtained in 2006. The inset shows the zoom of the time series during the flares. Shown is the delay on the flare event in the \ion{Ca}{ii} H and K lines and in \ion{He}{i} at 4026 \AA. Also shown are the pre-flare dips on the time series of these indicators.   }
    \label{fig:triangolo_flare}%
    \end{figure}

    Figure \ref{fig:triangolo_flare} shows the time series of normalised fluxes with respect to the quiescent state of the star. 
    By inspecting the time series of most activity indicators we can see the two points related to the flare where the flux decreases going from ID 79 to ID 80, two hours later, except for the \ion{He}{i} 4026 \AA \ and \ion{Ca}{ii} lines.
    
    In spite of the low temporal resolution we can see that \ion{He}{i} 4026 \AA \ and \ion{Ca}{ii} lines show a delay with respect to the other lines. 
    \citet{2006A&A...452..987C} reported a delay (up to 5 $\pm$ 3 minutes) for the \ion{Ca}{ii} and \ion{He}{i} 4026 \AA \ lines in some weak and short flares observed on AD Leonis. Our flare is much more intense and it is possible that this effect is enhanced with respect to the case of weaker flares. \citet{2003A&A...397.1019H} studied the dynamics of flares on dMe stars and show that the rise and decay times in the \ion{Ca}{ii} line are usually longer than the rise and decay times in the Balmer lines.

    \section{Luminosity and released energy}\label{appendice_energy}
    In order to estimate the flare energy released in the observed chromospheric lines, we have converted the observed flux to luminosity.
    The luminosity values obtained are provided in Table \ref{tab:luminositàenergiabrillamento} as $\mathrm{L_{max}}$ for the considered lines.
    We   calculated the value of luminosity for the quiescent state of star, $\mathrm{L_{quiet}}$, provided in Table \ref{tab:luminositàenergiabrillamento}, using the quiescent flux obtained from the average of the points outside the flare (red dashed lines in Fig. \ref{fig:triangolo_flare}). 
    Despite   the low temporal resolution we estimated the released energy by approximating the temporal evolution of the flare with a vertical ascent phase and a phase of linear decay. 
    
    \begin{figure}[hbp!]
    \centering
    \includegraphics[width=0.8\hsize]{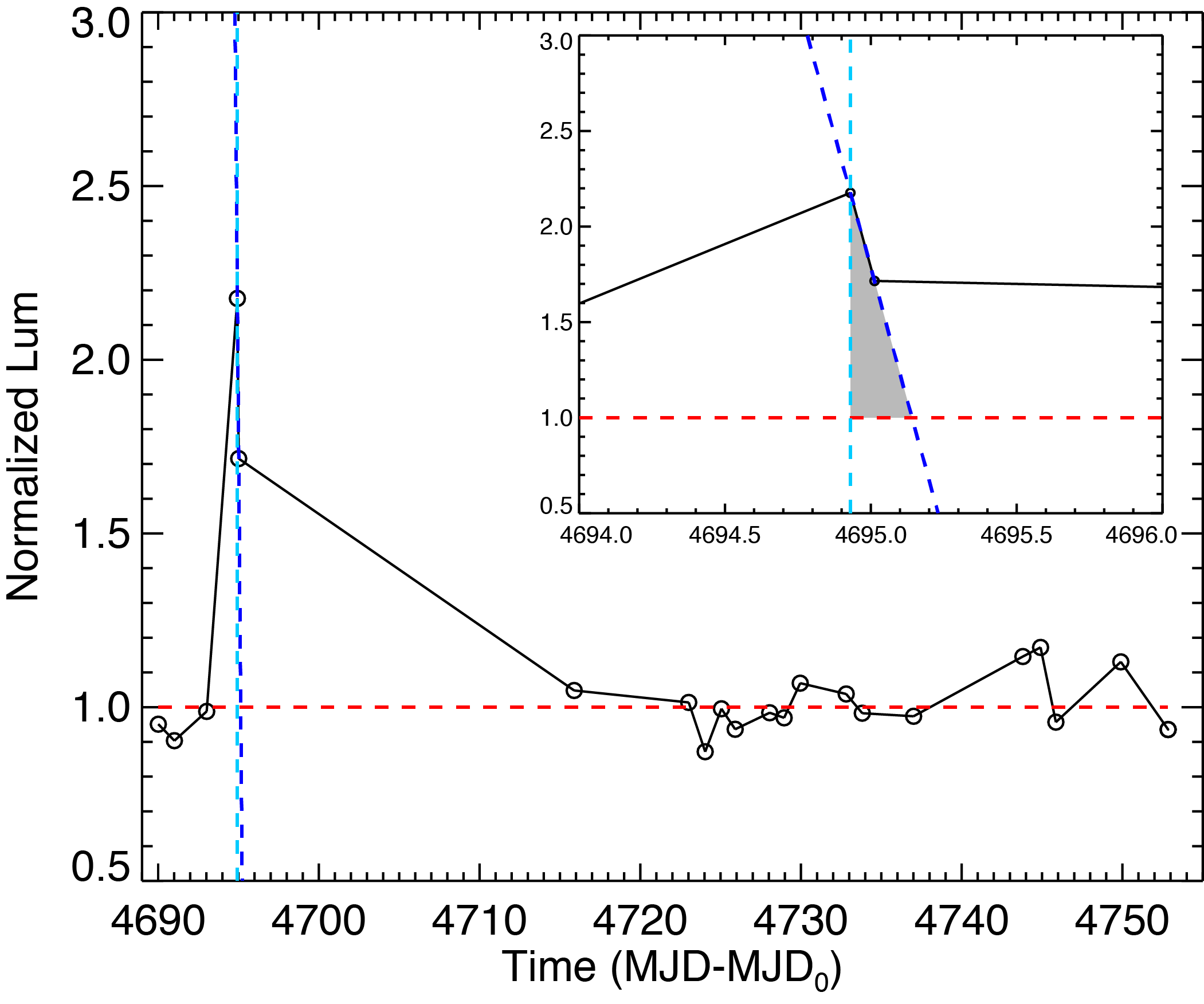}
    \caption{Example of the triangle used to calculate the energy released during the flare in a given line. The dashed light blue line, parallel to the ordinate axis and passing through the point of flare maximum, represents the rising phase of the flare; the dashed blue line, passing through the two points corresponding to the two observations performed two hours apart, approximate the decay phase of the flare. The dashed red line shows the quiescent state of the star. MJD$_0$ = 53758.244, time of the first observation obtained in 2006. The inset  shows the zoom-in on the triangle.} 
    \label{fig:triangolo_energia}%
    \end{figure}  
    
    Considering the characteristic timescale of the flares (from a few minutes to a few hours), it is unlikely that the flare ends in the next point, obtained with an observation carried out 20 days after the start of the flare. Therefore,  we have drawn a straight line passing through the two points corresponding to the two observations performed two hours apart (dashed  blue line) to reconstruct the shape of the flare (see Fig. \ref{fig:triangolo_energia}). The rising phase of the flare is approximated with one straight line (dashed  light blue line) parallel to the ordinate axis and passing through the point of flare maximum. From the area of this triangle we have obtained the value of energy released during the flare provided in Table \ref{tab:luminositàenergiabrillamento}. This is likely a conservative estimate since we cannot be sure that we observed the true flare maximum. 
    
    For the lines where the flare shows a delay we have only one point related to the flare, so we are not be able to calculate the energy released with the previous  technique mentioned. 
    
    \begin{table}[h]\small
        \begin{center}
                \caption{Value of maximum flux corresponding to the flare, value of luminosity for the quiescent state of the star and in correspondence of the maximum of the flare, and value of energy released during the flare. The errors take into account the error on the stellar radius, which has the greatest influence on the final values.}
                \label{tab:luminositàenergiabrillamento}
\begin{tabular}{lcccc}
            \toprule[0.05cm]
            \toprule
                        Line    &       F$_{max}$       &       L$_{quiet}$ & L$_{max}$ &   Energy  \\
                                &       \scriptsize{($10^5$ erg s$^{-1}$ cm$^{-2}$)}     &       \scriptsize{($10^{26}$ erg s$^{-1}$)}   & \scriptsize{($10^{26}$ erg s$^{-1}$)}&         \scriptsize{($10^{30}$ erg)}    \\
            \midrule
                        \ion{Ca}{ii} K  &       7.155   $\pm$   0.016   &       47      $\pm$   11      &       83      $\pm$   19      &       -                       \\
\ion{Ca}{ii} H  &       7.730   $\pm$   0.007   &       53      $\pm$   12      &       89      $\pm$   20      &       -                       \\
\ion{He}{i} 4026        &       0.282   $\pm$   0.012   &       1.5     $\pm$   0.3     &       3       $\pm$   0.7     &       -                       \\
\ion{He}{i} 4471        &       0.58    $\pm$   0.06    &       1.3     $\pm$   0.3     &       6.7     $\pm$   1.7     &       3.7     $\pm$   1.2     \\
H$\beta$        &       28.9    $\pm$   0.4     &       71      $\pm$   16      &       334     $\pm$   75      &       137     $\pm$   40      \\
\ion{He}{i} 5876        &       2.01    $\pm$   0.03    &       6.6     $\pm$   1.5     &       23      $\pm$   5       &       9       $\pm$   3       \\
\ion{Na}{i} D2  &       0.730   $\pm$   0.007   &       2.6     $\pm$   0.6     &       8.4     $\pm$   1.9     &       11      $\pm$   4       \\
\ion{Na}{i} D1  &       0.56    $\pm$   0.02    &       1.7     $\pm$   0.4     &       6.4     $\pm$   1.5     &       7       $\pm$   2       \\
H$\alpha$       &       23.86   $\pm$   0.13    &       127     $\pm$   28      &       276     $\pm$   62      &       136     $\pm$   62      \\
            \bottomrule[0.05cm]
        \end{tabular}
    \end{center}
    \end{table}
    
    We note that the luminosity of the maximum of the flare is greater than the luminosity of the quiescent state of the star by a factor of between $\sim 2$ and $\sim 5$. These results suggest that the ratio $\mathrm{L_{max}}/\mathrm{L_{quiet}}$ is higher for the lines we suppose are formed in the upper layers of the stellar atmosphere, like the helium lines.
    
    \onecolumn
    \section{Data}\label{appendice:tabelle}
    \begin{table*}[ht!]\scriptsize
        \begin{center}
                \caption{HARPS and HARPS-N observations.}
                \label{tab:ObsData}
                \begin{tabular}{cccc}
                        \toprule[0.05cm]
                        \toprule
                        \multicolumn{4}{c}{Dataset HARPS and HARPS-N}\\
                        \cmidrule(lr){1-4}
                         ID obs. & Date obs. & ID obs. & Date obs.  \\
                        \midrule
0       &       2006-01-23T05:50:40.777         &       49      &       2018-05-09T22:32:35.186         \\
1       &       2006-01-25T05:51:44.132         &       50      &       2018-05-09T23:47:46.114         \\
2       &       2006-01-26T06:28:08.689         &       51      &       2018-05-10T22:27:20.700         \\
3       &       2006-02-17T05:08:51.940         &       52      &       2018-05-10T23:39:34.923         \\
4       &       2006-02-19T05:10:02.902         &       53      &       2018-05-12T22:05:00.407         \\
5       &       2006-03-15T03:35:31.947         &       54      &       2018-05-12T23:45:11.480         \\
\end{tabular}

\end{center}
\end{table*}

    \begin{table*}\scriptsize
    \caption{}
    \label{tab:EW_2006}
    \centering
    \begin{tabular}{cccccccccc} 
    \toprule[0.05cm]
    \toprule
    \multicolumn{10}{c}{\textsc{Equivalent Width HARPS Data 2006}}\\
    \midrule[0.05cm]
    ID Obs. & EW$_{H\alpha}$ & EW$_{H\beta}$ & EW$_{CaII H}$ & EW$_{CaII K}$ & EW$_{CaII}$ & EW$_{HeI 4026}$ & EW$_{HeI 4471}$ & EW$_{HeI 5876}$ & EW$_{NaI}$\\
    & (\AA) & (\AA) & (\AA) & (\AA) & (\AA) & (\AA) & (\AA) & (\AA) & (\AA) \\
    \midrule
    0   &       3.40    $\pm$   0.09    &       3.43    $\pm$   0.10    &       10.86   $\pm$   0.09    &       18.92   $\pm$   0.17    &       12.47   $\pm$   0.08    &       0.407   $\pm$   0.004   &       0.126   $\pm$   0.016   &       0.059   $\pm$   0.017   &       0.36    $\pm$   0.03    \\
1       &       4.30    $\pm$   0.10    &       4.64    $\pm$   0.09    &       10.8    $\pm$   0.2     &       21.04   $\pm$   0.15    &       18.11   $\pm$   0.12    &       0.720   $\pm$   0.012   &       0.163   $\pm$   0.015   &       0.102   $\pm$   0.017   &       0.45    $\pm$   0.03    \\
2       &       3.45    $\pm$   0.09    &       3.60    $\pm$   0.11    &       10.74   $\pm$   0.04    &       18.6    $\pm$   0.2     &       10.91   $\pm$   0.04    &       0.439   $\pm$   0.013   &       0.126   $\pm$   0.017   &       0.069   $\pm$   0.018   &       0.36    $\pm$   0.02    \\
3       &       3.68    $\pm$   0.09    &       3.79    $\pm$   0.08    &       9.50    $\pm$   0.14    &       18.27   $\pm$   0.19    &       12.41   $\pm$   0.11    &       0.577   $\pm$   0.014   &       0.131   $\pm$   0.011   &       0.079   $\pm$   0.018   &       0.40    $\pm$   0.03    \\
4       &       3.27    $\pm$   0.09    &       3.24    $\pm$   0.11    &       11.13   $\pm$   0.06    &       18.24   $\pm$   0.08    &       13.76   $\pm$   0.05    &       0.489   $\pm$   0.011   &       0.103   $\pm$   0.015   &       0.045   $\pm$   0.017   &       0.33    $\pm$   0.03    \\
\end{tabular}
    \end{table*}
    
       \begin{table*}\scriptsize
    \caption{}
    \label{tab:EW_2018}
    \centering
    \begin{tabular}{cccccccccc} 
    \toprule[0.05cm]
    \toprule
    \multicolumn{10}{c}{\textsc{Equivalent Width HARPS-N Data 2018}}\\
    \midrule[0.05cm]
    ID Obs. & EW$_{H\alpha}$ & EW$_{H\beta}$ & EW$_{CaII H}$ & EW$_{CaII K}$ & EW$_{CaII}$ & EW$_{HeI 4026}$ & EW$_{HeI 4471}$ & EW$_{HeI 5876}$ & EW$_{NaI}$\\
    & (\AA) & (\AA) & (\AA) & (\AA) & (\AA) & (\AA) & (\AA) & (\AA) & (\AA) \\
    \midrule
33      &       5.23    $\pm$   0.10    &       5.67    $\pm$   0.15    &       12.95   $\pm$   0.06    &       20.2    $\pm$   0.3     &       13.29   $\pm$   0.06    &       0.898   $\pm$   0.013   &       0.190   $\pm$   0.014   &       0.135   $\pm$   0.017   &       0.59    $\pm$   0.03    \\
34      &       4.68    $\pm$   0.09    &       4.79    $\pm$   0.10    &       11.77   $\pm$   0.07    &       20.04   $\pm$   0.13    &       13.51   $\pm$   0.06    &       0.770   $\pm$   0.012   &       0.180   $\pm$   0.014   &       0.110   $\pm$   0.017   &       0.50    $\pm$   0.03    \\
35      &       5.12    $\pm$   0.10    &       4.99    $\pm$   0.06    &       8.7     $\pm$   0.5     &       19.0    $\pm$   1.5     &       9.5     $\pm$   0.4     &       0.663   $\pm$   0.017   &       0.20    $\pm$   0.06    &       0.20    $\pm$   0.02    &       0.60    $\pm$   0.02    \\
36      &       4.09    $\pm$   0.09    &       4.10    $\pm$   0.07    &       8.95    $\pm$   0.03    &       15.74   $\pm$   0.15    &       9.15    $\pm$   0.03    &       0.594   $\pm$   0.012   &       0.162   $\pm$   0.015   &       0.100   $\pm$   0.017   &       0.46    $\pm$   0.03    \\
37      &       4.32    $\pm$   0.10    &       4.63    $\pm$   0.09    &       11.63   $\pm$   0.12    &       17.10   $\pm$   0.10    &       14.87   $\pm$   0.08    &       0.760   $\pm$   0.011   &       0.174   $\pm$   0.014   &       0.100   $\pm$   0.017   &       0.48    $\pm$   0.03    \\
\end{tabular}
    \end{table*}
    
        \begin{table*}\small
    \caption{}
    \label{tab:Flx_2006_1}
    \centering
    \begin{tabular}{cccccc} 
    \toprule[0.05cm]
    \toprule
    \multicolumn{6}{c}{\textsc{Fluxes HARPS Data 2006}}\\
    \midrule[0.05cm]
    ID Obs. & Flx$_{H\alpha}$ & Flx$_{H\beta}$ & Flx$_{CaII H}$ & Flx$_{CaII K}$ & Flx$_{CaII}$ \\
    & ($10^5$ erg s$^{-1}$ cm$^{-2}$)& ($10^5$ erg s$^{-1}$ cm$^{-2}$)& ($10^5$ erg s$^{-1}$ cm$^{-2}$)& ($10^5$ erg s$^{-1}$ cm$^{-2}$)& ($10^5$ erg s$^{-1}$ cm$^{-2}$)\\
    \midrule
   0    &       9.24    $\pm$   0.19    &       5.03    $\pm$   0.09    &       4.817   $\pm$   0.008   &       4.905   $\pm$   0.009   &       4.857   $\pm$   0.006   \\
1       &       11.59   $\pm$   0.20    &       6.84    $\pm$   0.08    &       6.580   $\pm$   0.016   &       6.678   $\pm$   0.008   &       6.661   $\pm$   0.007   \\
2       &       9.32    $\pm$   0.19    &       5.33    $\pm$   0.10    &       5.503   $\pm$   0.007   &       5.668   $\pm$   0.012   &       5.545   $\pm$   0.006   \\
3       &       9.82    $\pm$   0.19    &       5.76    $\pm$   0.08    &       5.274   $\pm$   0.012   &       4.941   $\pm$   0.008   &       5.042   $\pm$   0.007   \\
4       &       8.79    $\pm$   0.19    &       4.78    $\pm$   0.11    &       5.163   $\pm$   0.007   &       5.035   $\pm$   0.007   &       5.098   $\pm$   0.005   \\
\end{tabular}
    \end{table*}

    \begin{table*}\small
    \caption{}
    \label{tab:Flx_2006_2}
    \centering
    \begin{tabular}{ccccc} 
    \toprule[0.05cm]
    \toprule
    \multicolumn{5}{c}{\textsc{Fluxes HARPS Data 2006}}\\
    \midrule[0.05cm]
    ID Obs. & Flx$_{HeI 4026}$ & Flx$_{HeI 4471}$ & Flx$_{HeI 5876}$ & Flx$_{NaI}$ \\
    & ($10^5$ erg s$^{-1}$ cm$^{-2}$)& ($10^5$ erg s$^{-1}$ cm$^{-2}$)& ($10^5$ erg s$^{-1}$ cm$^{-2}$)& ($10^5$ erg s$^{-1}$ cm$^{-2}$)\\
    \midrule
   0    &       0.099   $\pm$   0.012   &       0.073   $\pm$   0.012   &       0.44    $\pm$   0.03    &       0.131   $\pm$   0.003   \\
1       &       0.133   $\pm$   0.011   &       0.127   $\pm$   0.014   &       0.55    $\pm$   0.03    &       0.2686  $\pm$   0.0060  \\
2       &       0.099   $\pm$   0.013   &       0.086   $\pm$   0.013   &       0.44    $\pm$   0.03    &       0.160   $\pm$   0.007   \\
3       &       0.106   $\pm$   0.008   &       0.098   $\pm$   0.013   &       0.49    $\pm$   0.03    &       0.200   $\pm$   0.006   \\
4       &       0.080   $\pm$   0.011   &       0.055   $\pm$   0.013   &       0.41    $\pm$   0.03    &       0.171   $\pm$   0.006   \\
\end{tabular}
    \end{table*}

     \begin{table*}\small
    \caption{}
    \label{tab:Flx_2018_1}
    \centering
    \begin{tabular}{cccccc} 
    \toprule[0.05cm]
    \toprule
    \multicolumn{6}{c}{\textsc{Fluxes HARPS-N Data 2018}}\\
    \midrule[0.05cm]
    ID Obs. & Flx$_{H\alpha}$ & Flx$_{H\beta}$ & Flx$_{CaII H}$ & Flx$_{CaII K}$ & Flx$_{CaII}$ \\
    & ($10^5$ erg s$^{-1}$ cm$^{-2}$)& ($10^5$ erg s$^{-1}$ cm$^{-2}$)& ($10^5$ erg s$^{-1}$ cm$^{-2}$)& ($10^5$ erg s$^{-1}$ cm$^{-2}$)& ($10^5$ erg s$^{-1}$ cm$^{-2}$)\\
    \midrule
   33   &       14.04   $\pm$   0.19    &       8.58    $\pm$   0.11    &       5.836   $\pm$   0.007   &       4.717   $\pm$   0.009   &       5.466   $\pm$   0.005   \\
34      &       12.60   $\pm$   0.19    &       7.17    $\pm$   0.08    &       5.565   $\pm$   0.008   &       5.131   $\pm$   0.007   &       5.335   $\pm$   0.005   \\
35      &       13.78   $\pm$   0.18    &       8.97    $\pm$   0.05    &       3.62    $\pm$   0.03    &       3.03    $\pm$   0.03    &       3.324   $\pm$   0.018   \\
36      &       11.08   $\pm$   0.18    &       6.15    $\pm$   0.07    &       4.653   $\pm$   0.007   &       4.331   $\pm$   0.009   &       4.533   $\pm$   0.006   \\
37      &       11.7    $\pm$   0.2     &       6.84    $\pm$   0.08    &       4.770   $\pm$   0.009   &       4.224   $\pm$   0.007   &       4.418   $\pm$   0.005   \\
\end{tabular}
    \end{table*}

    \begin{table*}\small
    \caption{}
    \label{tab:Flx_2018_2}
    \centering
    \begin{tabular}{ccccc} 
    \toprule[0.05cm]
    \toprule
    \multicolumn{5}{c}{\textsc{Fluxes HARPS-N Data 2018}}\\
    \midrule[0.05cm]
    ID Obs. & Flx$_{HeI 4026}$ & Flx$_{HeI 4471}$ & Flx$_{HeI 5876}$ & Flx$_{NaI}$ \\
    & ($10^5$ erg s$^{-1}$ cm$^{-2}$)& ($10^5$ erg s$^{-1}$ cm$^{-2}$)& ($10^5$ erg s$^{-1}$ cm$^{-2}$)& ($10^5$ erg s$^{-1}$ cm$^{-2}$)\\
    \midrule
33      &       0.162   $\pm$   0.011   &       0.169   $\pm$   0.016   &       0.71    $\pm$   0.03    &       0.343   $\pm$   0.006   \\
34      &       0.149   $\pm$   0.010   &       0.135   $\pm$   0.014   &       0.60    $\pm$   0.03    &       0.292   $\pm$   0.006   \\
35      &       0.20    $\pm$   0.05    &       0.26    $\pm$   0.03    &       0.73    $\pm$   0.02    &       0.220   $\pm$   0.004   \\
36      &       0.135   $\pm$   0.012   &       0.126   $\pm$   0.014   &       0.56    $\pm$   0.03    &       0.226   $\pm$   0.006   \\
37      &       0.143   $\pm$   0.010   &       0.122   $\pm$   0.014   &       0.59    $\pm$   0.03    &       0.280   $\pm$   0.006   \\
\end{tabular}
    \end{table*}
    
   \end{appendix}
\end{document}